\documentclass[preprint,review,11pt]{elsarticle}

\usepackage{hyperref}
\usepackage{graphicx}
\usepackage{subfigure}
\usepackage{amssymb}	
\usepackage{mathtools}
\usepackage[numbers]{natbib}
\usepackage{bm}

\journal{Computer Methods in Applied Mechanics and Engineering}

\begin{document}
\begin{frontmatter}

\title{Computational Model for Granular Flows based on the Lagrangian Particle Method}

\author[1]{Mario Zepeda}

\author[1,2]{Roman Samulyak\corref{Roman Samulyak}}
\cortext[Roman Samulyak]{Corresponding author}
\ead{roman.samulyak@stonybrook.edu}

\address[1]{Department of Applied Mathematics and Statistics, Stony Brook University, Stony Brook, New York 11794, USA.}
\address[2]{Computational Science Initiative, Brookhaven National Laboratory, Upton, New York 11973, USA.}

\begin{abstract}
A numerical model and parallel software for 3D simulations of granular flows have been developed based on the Lagrangian particle (LP) method [R. Samulyak, X. Wang, H.-C. Chen, Lagrangian particle method for compressible fluid dynamics, J. Comput. Phys. 362 (2018) 1–19], originally developed for compressible hydrodynamic flows, including free surface / multiphase problems. LP uses local polynomial least square fittings on particle-based stencils that ensure numerical convergence to the prescribed order.  The granular flow model implements continuum equations with a $\mu(I)$-rheology closure that is capable of describing two-directional transitions of the flow between various regimes characterized by solid-, liquid-, and gas-like features. The granular flow code has been parallelized for distributed memory supercomputers and validated by comparing 3D simulations to experimental data on the collapse of granular columns. Numerical simulations showing the granular flow transition to a gas-like regime have also been presented.

\end{abstract}

\end{frontmatter}

\section{Introduction}

A continuum description of granular flow is of fundamental importance for industrial applications and ongoing scientific research. Industrial applications are typical in agriculture, geophysics, and the pharmaceutical sector, to name a few. In geophysical sciences, simulations are particularly useful for the prediction of natural phenomena such as landslides or avalanches which could represent a hazard to human life \cite{avalanche}. They are also important in industrial applications that deal with the transportation of solid material in powder or grain through channels and narrow passages where the design of bulk material handling equipment requires a deep understanding of the mechanical properties of the granular material \cite{chuteIndustry}. Another class of applications use motion to separate grains of varied sizes and properties \cite{Industry2}. Granular flow also has applications in the pharmaceutical industry where granulation i.e., the process of producing grains or granules from a solid substance or a powder is commonly used and the compression properties of materials need to be understood for creating pills and solvable products \cite{granulation}. 

Granular flows are also subject of scientific research. One example is the study of tungsten powder interacting with high energy proton beams \cite{tungstenBeam}. It is expected that tungsten powder, as a high-atomic number material, may be useful as a target material for future high energy particle accelerators. The purpose of targets is to convert relativistic particle beams that can be easily obtained and accelerated, such as proton beams, into beams of other elementary articles such as pions or neutrons \cite{Abrams2011}. The tungsten powder will be able to withstand great thermal loads imposed by particle beams. New advanced simulation tools are needed to fully explore the feasibility of granular tungsten targets for future accelerators.

The study of granular flow has been an active area of research over last decades, resulting in the development of methods that are very different in numerical approximations and applicability range. The Discrete Element Method (DEM) \cite{DEM1,DEM2,DEM3} computes direct interactions between grains. While DEM is gaining more popularity with the increase of computing hardware capabilities, it is still computationally impracticable for large systems and long times.

In the continuum modeling framework, difficulties arise because the granular media can change its behavior between different flow regimes. Granular flow studies typically outline the following three regimes \cite{Chialvo2012}:
\begin{enumerate}	
	\item In the \textbf{quasi-static regime} the grain inertia is negligible; this regime is usually described using solid plasticity models.
	\item In the \textbf{gas-like regime} the grains are strongly agitated and usually far apart of each other. This regime has been modeled by analogy with the kinetic theory of gases in which the grains interact by binary collisions.
	\item By far the most controversial of the three regimes is the one in which the grains behave like a dense liquid, usually called \textbf{inertial regime}. In this dense flow regime, many different constitutive models have been proposed but there no general consensus has yet been reached.
\end{enumerate}

As of today, there are no constitutive equations capable of describing the dynamics of a physical process that changes through the different regimes. However, a variety of closure equations using plasticity theory  \cite{SoilPlasticity1,Soilplasticty2,SoilPlasticity3} and using kinetic theory of granular flow \cite{kinetic1,kinetic2} produce acceptable results within the quasi-static regime and the gas-like regime respectively.  As for the inertial regime, the $\mu(I)$-rheology proposed in \cite{gdr2004dense, Jop2006Jun} lead to a progress in obtaining closure equations in \cite{lagree2011granular} and more recently in  \cite{Barker2017, Schaeffer2019Sep}, where the inertial compressible $I$-dependent rheology was introduced.

With any appropriate choice of the closure model, simulations of granular flows based on continuous equations may use the conventional grid-based approaches or other approximations. Some features of granular flows such as large deformations and complex free surfaces create difficulties for using mesh-based methods. Material Point Method (MPM) and Smooth Particle Hydrodynamics (SPH) methods better handle transitions through dense and disconnected states as shown in \cite{Dunatunga2015Sep,MPM1} and \cite{Yang2020May,SPH2}. However, it is  known that the discretization of spatial derivative in the SPH method suffers from accuracy issues. It is widely accepted \cite{Monaghan2005,Hopkins2014} that the traditional SPH discretization has zeroth-order
convergence for widely used kernels and numerous methods have been developed to improve the SPH discretization of fluid dynamics equations.

In this paper, we proposed a 3D numerical model and parallel software for granular flows based on the Lagrangian particle (LP) method for compressible hydrodynamics \cite{Samulyak2018Jun}. In the LP method, approximations of spatial derivatives are obtained by employing a local polynomial fitting known also as the generalized finite difference (GFD) method \cite{BenitoUrena2001}, which is always accurate to a prescribed order. The LP method for compressible hydrodynamics is second-order accurate in space and time and it is extendable to higher order accuracy. In addition to compressible hydrodynamics, the LP code implements equations of low magnetic Reynolds number magnetohydrodynamics \cite{Samulyak2021Apr}. Our granular flow model uses a $\mu(I)$-rheology model for the inertial flow regime, a linear pressure-density relation in the quasi-static regime, and it supports the transition to a gas-like regime when the density decreases below a minimum value for the inertial regime. 

The paper is organized as follows. In Section \ref{MathModel}, the governing system of equations (conservation laws) is described followed by the  $\mu(I)$-rheology model and the description of transitions between various flow regimes. Numerical implementation of the granular flow model using the Lagrangian particle framework is described in Section \ref{NSaI}. The main ideas of parallelization of particle data structures and particle handling for massively parallel supercomputers and scalability tests are also presented.  In Section \ref{VandV}, we report results of the code validation by comparing numerical simulations with experimental data on the collapse of a granular column, followed by a demonstration of the code capabilities to handle transitions between flow regimes. Finally, we conclude this paper with a summary of our results and plans for the future work.

\section{Mathematical model for granular flows}\label{MathModel}

\subsection{Conservation law equations}

In continuum modeling, granular flows are described by the solid fraction (also known as volume fraction) $\phi$, the velocity vector $ \bm u = (u,v,w)$, and the symmetric Cauchy stress tensor $ \bm \sigma = (\sigma_{ij})$. 
These quantities are related by the mass and momentum conservation equations in the Lagrangian formulation as: 
\begin{eqnarray} 
\phi_t &=&  -\phi \left( \frac{ \partial u}{ \partial x } + \frac{ \partial v}{ \partial y} + \frac{ \partial w}{ \partial z} \right),\\
u_t &=&  \frac{1}{\phi_s  \phi} \left(  \frac{ \partial \sigma_{xx}}{ \partial x } + \frac{ \partial \sigma_{xy}}{ \partial y} + \frac{ \partial \sigma_{xz}}{ \partial z} \right) + g_x \label{eq:sys2},\\
v_t &=&  \frac{1}{\phi_s  \phi} \left( \frac{ \partial \sigma_{yx}}{ \partial x } + \frac{ \partial \sigma_{yy}}{ \partial y} + \frac{ \partial \sigma_{yz}}{ \partial z} \right)  +  g_y,\\
w_t &=& \frac{1}{\phi_s  \phi} \left( \frac{ \partial \sigma_{zx}}{ \partial x } + \frac{ \partial \sigma_{zy}}{ \partial y} + \frac{ \partial \sigma_{zz}}{ \partial z} \right)   +  g_z,\label{eq:sys4}
\end{eqnarray}
where $\phi_s$ is the constant material density of grains and $g_{\alpha}$ represents an external force. 
In order to obtain expressions for the Cauchy stress tensor $\sigma$, we start by decomposing it into a diagonal pressure term and a trace-free tensor $\bm \tau$:
\begin{equation}
\sigma_{ij} = -P \delta_{ij} + \tau_{ij}.
\end{equation}
The tensor $\bm \tau$ is called the shear stress tensor or the deviatoric stress tensor.

To close the system, a formulation for the pressure and the deviatoric stress tensor in terms of velocity and density is required. This formulation will depend on the flow regime. For the inertial regime, the $\mu(I)$-rheology constitutive model is presented in the next section.

\subsection{Constitutive relations} \label{ConRel}

\begin{figure}[h!]
	\centering
	\includegraphics[width=0.4\linewidth]{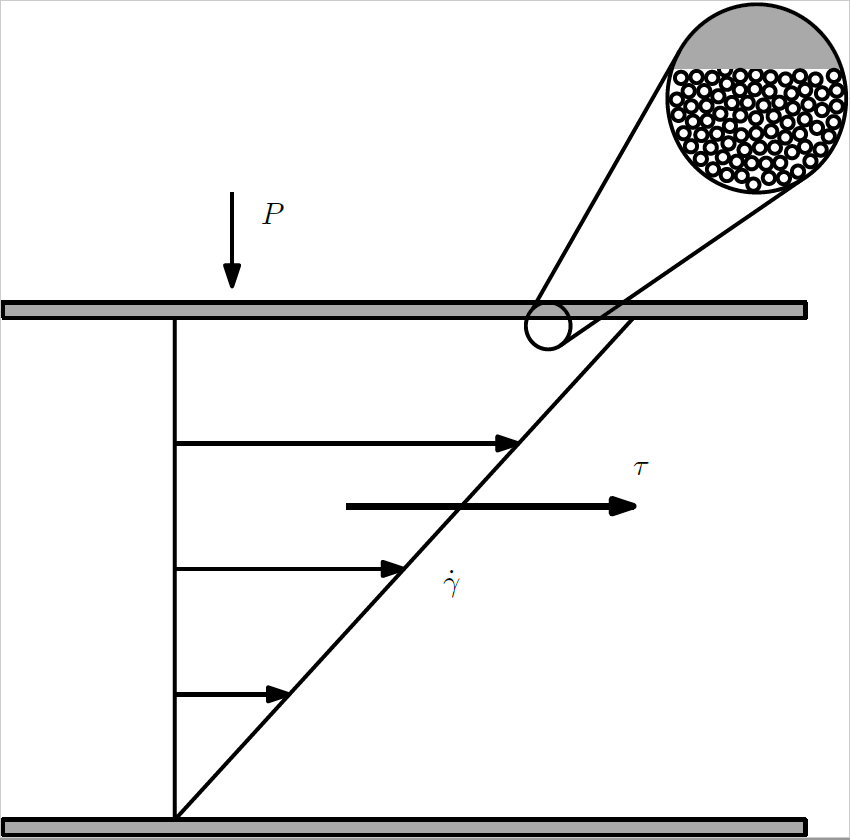}
	\caption{Schematic of normal pressure $P$, shear stress $\bm \tau$ and shear rate $\bm {\dot \gamma}$ in the simple plane shear configuration.}
	\label{fig:dia}
\end{figure}

A constitutive model widely used for the inertial regime, called $\mu(I)$-rheology \cite{Jop2006Jun}, considers the following configuration: a granular material is confined under a normal stress $P$ between two rough planes, undergoing a rate $\bm {\dot\gamma}$ shear by a shear stress $ \bm \tau$, as shown in figure \ref{fig:dia}.

The shear rate tensor $\bm {\dot \gamma}$ is completely determined in terms of velocity as
\begin{equation}
\dot \gamma_{ij} = D_{ij} - \frac{1}{3} \text{div}(\bm u) \delta_{ij},
\end{equation}
where $\bm D$ is the strain rate tensor given by:
\begin{equation} \label{eqn:strainrate}
D_{ij} = \frac{1}{2} \left(  \frac{ \partial u_i}{\partial x_j} + \frac{ \partial u_j}{\partial x_i } \right). 
\end{equation}

It was shown in \cite{daCruz2005Mar} and \cite{Lois2005Nov} that the system can be described using a dimensionless parameter called the inertial number $I$. The inertial number can be interpreted in terms of the macroscopic time scale and a inertial time scale.
Consider the motion of one grain during a simple shear as shown in figure \ref{fig:dia}. The grain obeys the deformation given by $ \left|  \bm{\dot \gamma} \right| $ ocurring in time
\[
T_{macro} = \frac{1}{2 \left|  \bm{\dot \gamma} \right|}
\]
until it passes over a void region; at that moment the grain is pushed downward due to the confining pressure. The time it takes for this rearrangement to occur can be estimated by the free fall of the grain of diameter $d$ and density $\phi_s$ under a force $Pd^2$ across a distance $d$:
\[
T_{inertial} = \frac{d}{\sqrt {P/\phi_s}}
\]
The inertial number is defined as the ratio of the inertial time scale and the macroscopic time scale as
\begin{equation} \label{eqn:inertial}
I = \frac{T_{inertial}}{T_{macro}} = \frac{2 \left| \bm{\dot \gamma} \right|  d}{\sqrt{P/\phi_s}}. 
\end{equation}

Using dimensional arguments, it can be shown that the shear stress $\tau$ is proportional to the pressure $P$, with a coefficient of proportionality $\mu(I)$ which depends only on $I$: 
\begin{equation}
\left|  \bm \tau \right|  = \mu(I) P,
\label{eqn:friction}
\end{equation} 
where $\mu(I)$ is the effective friction coefficient. This relation is usually referred as the yield condition or \textbf{friction constitutive equation} for $\mu(I)$-rheology. The friction constitutive equation implicitly expresses the idea that a granular material cannot deform unless the shear stress is sufficient to overcome friction. 
It was observed in \cite{Pouliquen1999Mar} and \cite{Pouliquen2002Feb} that $\mu (I)$ can be described as a rational function 
\begin{equation}
\mu(I) = \mu_s + \frac{\mu_2 - \mu_s}{I_0/I + 1}
\end{equation}
ranging from value $\mu_s$ at zero shear rate and converging asymptotically to a limit value $\mu_2$ for large $I$. 
The parameter $\mu_s$ corresponds to the tangent of the static angle of repose. This material dependent angle is the slope at which flow stops for certain granular material. The coefficient $\mu_2$ corresponds to the tangent of the angle of motion which is the critical angle at which the material starts moving in a granular heap.   
$I_0$ is a constant that scales the inertial number  and depends on the flow layer thickness, the depth-averaged velocity and some material-dependent constants \cite{jop_forterre_pouliquen_2005}.

The volume fraction is also a function of the inertial number, therefore there is a function $\Phi$ such that
\begin{equation}
\phi = \Phi(I).
\end{equation}
This function implicitly gives a \textbf{constitutive equation} relating the inertial number and the solid fraction. Experiments and numerical simulations have shown that this function is a decreasing function of $I$. The time-averaged quantity is gives \cite{Pouliquen_2006A} as 
\begin{equation} 
\phi = \frac{T_{inertial}\phi_{min}  + (T_{macro} - T_{inertial})\phi_c }{T_{macro}},
\end{equation}
that is equivalent to 
\begin{equation} \label{eqn:inertialdensity} 
\phi = \phi_c - ( \phi_c - \phi_{min})I.
\end{equation}
Here $\phi_c$ is the critical density or the upper bound for the solid fraction state and $\phi_{min}$ is the lower bound or the solid fraction density at which grains lose contact among themselves. 

Defining $a = \phi_c - \phi_{min}$ and using equations (\ref{eqn:inertial}) and (\ref{eqn:inertialdensity}), we describe the pressure as a function of the solid fraction and the norm of the shear rate tensor:
\begin{equation}\label{muofphi}
P(\phi) = \phi_s \left(  \frac{ 2a d  \left| \bm{\dot \gamma} \right|  }{\phi_c - \phi} \right)^2.
\end{equation} 
Note that the pressure depends on the square of the shear rate tensor and the grain diameter, consistent with Bagnold scaling \cite{Bagnold}.

To obtain an expression for the shear stress tensor, it is assumed that the deviatoric stress tensor satisfies the \textbf{alignment constitutive equation} \cite{Barker2017}  which takes the form:\
\begin{equation} \label{eqn:align}
\frac{\tau_{ij}}{\left|  \bm \tau  \right| } = \frac{ \dot \gamma_{ij}}{\left|  \bm{\dot \gamma} \right|  }. 
\end{equation}
Using equations (\ref{eqn:friction}) and (\ref{eqn:align}), we obtain the shear stress tensor as 
\begin{equation} \label{eqn:CST1}
\tau_{ij} = \eta(\left| \dot \gamma \right| ,P) \dot \gamma_{ij}  =\frac{\mu(I) P }{ \left| \bm{\dot \gamma} \right|  } \dot \gamma_{ij},
\end{equation}
where $\eta$ is an effective viscosity which goes to infinity when the shear rate goes to zero. Therefore, a yield criterion mush exist and the material starts flowing only if the following condition is satisfied
\begin{equation} \label{eqn:ycriterion}
 \left| \tau \right|  > \mu_s P. 
\end{equation}

\subsubsection{Transition to other flow regimes}

In this work, we propose a model capable of managing transitions between the quasi-static regime, the inertial regime, and the gas-like regime. The model has been implemented in 3D, which is, to the best of our knowledge, the first implementation of this kind.

As the granular material approaches the critical density, the inertial regime constitutive equations no longer describes the dynamics of the process correctly. 
In the inertial regime, the pressure depends on the density and the shear rate tensor, however in this quasi-static regime, the pressure no longer depends on the shear rate tensor as noted in \cite{Chialvo2012}.
Once the density reaches a maximum density $\phi_{max}$, the material is considered to be in the quasi-static regime and the following relation for pressure is proposed:
\begin{equation} \label{eqn:CST2}
P = \phi_s (\alpha + \beta \phi),
\end{equation}
where $\alpha$ and $\beta$ are constants depending on rheological parameters. Similarly, the Cauchy stress tensor depends only on the isotropic pressure in the form
\begin{equation} \label{eqn:CST3}
\sigma_{ij} = \phi_s (\alpha + \beta \phi) \delta_{ij}.
\end{equation}
In the case of rarefaction, the pressure approaches zero as grains begin to lose contacts and the granular material behaves somewhat similar to a gas phase. Numerically, we assign zero pressure to granular particles in this regime and their motion is controlled by inertia.

The pressure relation for the combined regimes is as follows:
\begin{equation} \label{eqn:EOS2}
P(\phi) = 
\begin{cases}
0  &\quad\text{if } \phi \leq \phi_{min}, \\
\phi_s \left(  \frac{ 2a d  \left| \bm{\dot \gamma} \right|  }{\phi_c - \phi} \right)^2  &  \quad\text{if }  \phi_{min} <   \phi \leq \phi_{max}, \\
\phi_s (\alpha + \beta \phi) & \quad\text{if }  \phi_{max} <   \phi \leq \phi_{c}, \\
\end{cases}
\end{equation}
where $\phi_{min}$ denotes the minimum density at which grains lose contact and no longer interact with each other.
We believe our model gives the first full 3D implementation of the $\mu(I)$-rheology model with a pressure expression based on rheological considerations. In \cite{Yang2020May}, a 3D implementation based in SPH is presented, however the equation for pressure in the inertial regime is empirical. 

\subsection{Combined system of equations}

In this Section, we combine expressions introduced above and obtain equations suitable for numerical implementation.
Introducing two supplementary scalar functions that depend only on density as
\begin{eqnarray}
M(\phi) =  \frac{ - 4 a^2 d^2}{(\phi_c - \phi)^2}, \\
N(\phi) =    \frac{ 4 a^2 d^2}{(\phi_c - \phi)^2} \frac{a \mu_s I_0 + \mu_2(\phi_c - \phi) }{aI_0 + (\phi_c - \phi)},   
\end{eqnarray}
the Cauchy stress tensor takes the  following form:
\begin{equation*}
\frac{1}{\phi_s} \sigma_{ij} = M(\phi) \left| \bm{\dot \gamma} \right|^2  \delta_{ij} + N(\phi) \left| \bm{\dot \gamma} \right|  \dot \gamma_{ij} 
\end{equation*}
Similarly, we define functions $S(\phi)$ and $R(\phi)$ that simplify expressions for partial derivatives of $M(\phi)$ and $N(\phi)$, respectively:
\begin{eqnarray*}
	\frac{\partial M(\phi)}{\partial x_j} = \frac{-8a^2 d^2}{\left(\phi_c - \phi\right)^3}\frac{\partial \phi}{\partial x_j} = S\left(\phi\right)\frac{\partial \phi}{\partial x_j}, \\
	\frac{\partial N(\phi)}{\partial x_j} = \left[ -S(\phi) \frac{a \mu_s I_0 + \mu_2 \left(\phi_c - \phi\right)}{ aI_0 + \phi_c - \phi}  - M(\phi)\frac{aI_0\left(\mu_2 - \mu_s\right)}{\left(aI_0 + \phi_c - \phi\right)^2} \right]  \frac{\partial \phi}{\partial x_j} = R\left(\phi\right)\frac{\partial \phi}{\partial x_j}.
\end{eqnarray*}

For computational purposes, we transform $\phi$ to volume as
\[
\frac{\partial \phi}{\partial x_j} = \frac{\partial}{\partial x_j}\left(\frac{1}{V}\right) = -\frac{1}{V^2}\frac{\partial V}{\partial x_j}.
\]

Combining all necessary expressions, we obtain the following system of partial differential equations:
\begin{equation*}
V_t = V\left(\text{div}\left(\bm u\right)\right),
\end{equation*}
\begin{equation*}
\begin{split}
u_t =& V \left[  \frac{- R\left(V\right)\left|\dot{\gamma}\right|}{V^2}\left(\left(\frac{S\left(V\right)}{R\left(V\right)}\left|\dot{\gamma}\right| + \dot{\gamma}_{11}\right)\frac{\partial V}{\partial x} + \dot{\gamma}_{12}\frac{\partial V}{\partial y} + \dot{\gamma}_{13}\frac{\partial V}{\partial z}\right) \right.  \\
+& N\left(V\right)\left(\left(\frac{2M\left(V\right)}{N\left(V\right)}\left|\dot{\gamma}\right| + \dot{\gamma}_{11}\right)\frac{\partial \left|\dot{\gamma}\right|}{\partial x} + \dot{\gamma}_{12}\frac{\partial \left|\dot{\gamma}\right|}{\partial y} + \dot{\gamma}_{13}\frac{\partial \left|\dot{\gamma}\right|}{\partial z}\right) \\
+& \left.  N\left(V\right)\left|\dot{\gamma}\right|\left(\frac{1}{2}\Delta u + \frac{1}{6}\frac{\partial}{\partial x}\left(\text{div}\left(\bm u\right)\right)\right)     \right] ,
\end{split}
\end{equation*}

\begin{equation*}
\begin{split}
v_t  =& V \left[ \frac{- R\left(V\right)\left|\dot{\gamma}\right|}{V^2} \left(\dot{\gamma}_{21}\frac{\partial V}{\partial x} + \left(\frac{S\left(V\right)}{R\left(V\right)}\left|\dot{\gamma}\right| + \dot{\gamma}_{22}\right)\frac{\partial V}{\partial y} + \dot{\gamma}_{23}\frac{\partial V}{\partial z}\right) \right. \\ 
+& N\left(V\right)\left(\dot{\gamma}_{21}\frac{\partial \left|\dot{\gamma}\right|}{\partial x} + \left(\frac{2M\left(V\right)}{N\left(V\right)}\left|\dot{\gamma}\right| + \dot{\gamma}_{22}\right)\frac{\partial \left|\dot{\gamma}\right|}{\partial y} + \dot{\gamma}_{23}\frac{\partial \left|\dot{\gamma}\right|}{\partial z}\right) \\
+& \left. N\left(V\right)\left|\dot{\gamma}\right|\left(\frac{1}{2}\Delta v + \frac{1}{6}\frac{\partial}{\partial y}\left(\text{div}\left(\bm u\right)\right)\right) \right] - g,
\end{split}
\end{equation*}

\begin{equation*}
\begin{split}
w_t  =& V \left[ \frac{- R\left(V\right)\left|\dot{\gamma}\right|}{V^2} \left(\dot{\gamma}_{31}\frac{\partial V}{\partial x} + \dot{\gamma}_{32}\frac{\partial V}{\partial y} + \left(\frac{S\left(V\right)}{R\left(V\right)}\left|\dot{\gamma}\right| + \dot{\gamma}_{33}\right)\frac{\partial V}{\partial z}\right) \right. \\ 
+& N\left(V\right)\left(\dot{\gamma}_{31}\frac{\partial \left|\dot{\gamma}\right|}{\partial x} + \dot{\gamma}_{32}\frac{\partial \left|\dot{\gamma}\right|}{\partial y} + \left(\dot{\gamma}_{33} + \frac{2M\left(V\right)}{N\left(V\right)}\left|\dot{\gamma}\right|\right)\frac{\partial \left|\dot{\gamma}\right|}{\partial z}\right) \\
+& \left. N\left(V\right)\left|\dot{\gamma}\right|\left(\frac{1}{2}\Delta w + \frac{1}{6}\frac{\partial}{\partial z}\left(\text{div}\left(\bm u\right)\right)\right) \right] .
\end{split}
\end{equation*}


\section{Numerical implementation}\label{NSaI}

\subsection{Outline of main time step algorithms}

The Lagrangian Particle (LP) for solving Euler equations for compressible fluids \cite{Samulyak2018Jun} draws some inspiration from the smooth particle hydrodynamics method (SPH) method \cite{Monaghan2005}. Unlike SPH, it approximates spatial derivatives by employing a local polynomial fitting, known also as the generalized finite difference (GFD) method \cite{BenitoUrena2001}, which is always accurate to a prescribed order. This it corrects the numerical accuracy issue of SPH described in the Introduction. The LP method compressible hydrodynamics is second-order accurate
in space and time and it is extendable to higher order accuracy. An outline of the Lagrangian Particle numerical method to solve the equations of granular flow is presented below. 
\begin{enumerate}
	\item \textbf{Initialization Setup.} An input file with the problem data (geometry, material properties, numerical resolution given by  the interparticle distance, initial time, final time, type of boundary conditions etc.) is read, initial data structures are created, including the octree structure for particle storage. 
	
	\item \textbf{Creation of Initial Particle Distribution.} An initial distribution of Lagrangian particles is created based on the problem geometry and numerical resolution.  
	
	\item \textbf{Creation of Boundary Particles.} Special particles designed to model solid boundaries or inflow / outflow boundary conditions are created.  
	
	\item \textbf{Octree.} An octree if constructed for the present distribution of particles representing granular material and boundaries. THe algorithm adaptively refines or coarsens the octree until the number of particles in an octant does not exceed a prescribed number.
	
	\item \textbf{Neighbor searching.} The routine finds  neighbors for each particles in all directions, in particular, in all neighboring octants. If the free boundary condition is applied and neighbors are not complete in certain directions, ghost particles are generated to complete the neighborhood.

	\item \textbf{Stencil Creation.} Given the neighbors of each particles, numerical stencils are generated. Stencils are close to spherical in shape for the granular flow solver. A simple sorting of particles in the distance ascending order may lead to the selection of most particles from one direction while missing information in other directions. Thus, the second principle of ordering neighbor lists is that the top elements (particles) in each list should be located in all directions (in all neighboring octants).

	\item \textbf{Approximate Differential Operators.} The generalized finite difference / local polynomial fitting \cite{BenitoUrena2001} is used to compute numerical approximations for spatial derivatives. To calculate the derivatives of state $s$ at the location of particle 0 using neighboring particles denoted by index $i$, consider the following truncated  Taylor series expansion (2D example was chosen for simplicity):
	\begin{equation*}
	s_i = s_0 + h_i \left.\frac{\partial s}{\partial x}\right|_0 + k_i \left.\frac{\partial s}{\partial y}\right|_0 + \frac{1}{2} \left( h^2_i \left.\frac{\partial^2 s}{\partial x^2}\right|_0 + k^2_i \left.\frac{\partial^2 s}{\partial y^2}\right|_0  + 2 h_i k_i \left.\frac{\partial^2 s}{\partial x \partial y}\right|_0 \right), 
	\end{equation*}
	where $h_i$ and $k_i$ denote the distance from particle $i$ to particle 0 in $x$ and $y$ directions, respectively.
	If this expansion is performed for $m$ particles in the neighborhood of particle 0, the unknown derivatives can be found by solving the following least-squares problem: 
	\[
	\begin{pmatrix}
	h_1 & k_1  & \frac{1}{2}h_1^2 & \frac{1}{2}k_1^2  & h_1k_1\\
	h_2 & k_2  & \frac{1}{2}h_2^2 & \frac{1}{2}k_2^2  & h_2k_2\\
	\vdots{} & \vdots{} & \vdots{} & \vdots{} & \vdots{} \\
	h_m & k_m  & \frac{1}{2}h_m^2 & \frac{1}{2}k_m^2  & h_mk_m\\
	\end{pmatrix}
	\begin{pmatrix}
	s_x \\
	s_y\\
	s_{xx} \\
	s_{yy}\\
	s_{xy}
	\end{pmatrix}
	= 
	\begin{pmatrix}
	s_1 - s_0 \\
	s_2 - s_0\\
	\vdots{} \\
	s_m - s_0
	\end{pmatrix}
	\]
	The system $Ax = b$ is usually over determined. An optimal solution, found via the QR decomposition, minimizes $\left\| Ax - b\right\|_2$.

	\item \textbf{Time Integration.}
	The difference equations are explicitly given by:
	\begin{equation*}
		\frac{V^{n+1} - V^n}{\Delta t} = V\left(\text{DIV}\left(\bm u\right)\right),
	\end{equation*}
		
	\begin{equation*}
		\begin{split}
			\frac{u^{n+1} - u^n}{\Delta t} =& V \left[  \frac{- R\left|\dot{\gamma}\right|}{V^2}\left(\left(\frac{S}{R}\left|\dot{\gamma}\right| + \dot{\gamma}_{11}\right) V_X + \dot{\gamma}_{12} V_Y + \dot{\gamma}_{13} V_Z \right) \right.  \\
			+& N \left(\left(\frac{2M}{N}\left|\dot{\gamma}\right| + \dot{\gamma}_{11}\right)  \left|\dot{\gamma}\right|_X  + \dot{\gamma}_{12} \left|\dot{\gamma}\right|_Y + \dot{\gamma}_{13} \left|\dot{\gamma}\right|_Z \right) \\
			+& \left.  N \left|\dot{\gamma}\right|\left(\frac{1}{2} \text{LAP} \left(u \right) + \frac{1}{6} \left(\text{DIV}\left(\bm u\right)\right)_X \right)     \right],
		\end{split}
	\end{equation*}
	
	\begin{equation*}
		\begin{split}
			\frac{v^{n+1} - v^n}{\Delta t}  =& V \left[ \frac{- 
				R \left|\dot{\gamma}\right|}{V^2} \left(\dot{\gamma}_{21} V_X + \left(\frac{S}{R}\left|\dot{\gamma}\right| + \dot{\gamma}_{22}\right) V_Y + \dot{\gamma}_{23} V_Z \right) \right. \\ 
			+& N \left(\dot{\gamma}_{21} \left|\dot{\gamma}\right|_X + \left(\frac{2M}{N}\left|\dot{\gamma}\right| + \dot{\gamma}_{22} \right) \left|\dot{\gamma}\right|_Y + \dot{\gamma}_{23} \left|\dot{\gamma}\right|_Z \right) \\
			+& \left.  N \left|\dot{\gamma}\right|\left(\frac{1}{2} \text{LAP} \left(v \right) + \frac{1}{6} \left(\text{DIV}\left(\bm u\right)\right)_Y \right)     \right] - g,
		\end{split}
	\end{equation*}
	
	\begin{equation*}
		\begin{split}
			\frac{w^{n+1} - w^n}{\Delta t}  =& V \left[ \frac{- 
				R\left|\dot{\gamma}\right|}{V^2} \left(\dot{\gamma}_{31} V_X + \dot{\gamma}_{32} V_Y + \left(\frac{S}{R}\left|\dot{\gamma}\right| + \dot{\gamma}_{33}\right) V_Z \right) \right. \\ 
			+& N\left(\dot{\gamma}_{31} \left|\dot{\gamma}\right|_X + \dot{\gamma}_{32} \left|\dot{\gamma}\right|_Y + \left(\dot{\gamma}_{33} + \frac{2M}{N}\left|\dot{\gamma}\right|\right) \left|\dot{\gamma}\right|_Z \right) \\
			+& \left.  N \left|\dot{\gamma}\right|\left(\frac{1}{2} \text{LAP} \left(w \right) + \frac{1}{6} \left(\text{DIV}\left(\bm u\right)\right)_Z \right) \right].
		\end{split}
	\end{equation*}
	Here DIV and LAP are the second-order accurate approximation of the divergence and Laplacian operator, respectively, using (approximately) symmetric stencils (see the description of stencils above). The sub-indexes  X,Y,Z denote partial derivatives computed using the GFD approximations. The spatial derivatives are second order accurate, but the time derivatives are only first order accurate. However, since a typical granular flow simulation is typically very stiff and may go through changes of regimes (gas-like, inertial, quasi-static), it is necessary to take small time steps. Therefore, the second-order accuracy in time is not as critical as it is in hydrodynamic systems. Nevertheless, it is planned to implement a second order accurate temporal discretization scheme that would improve the overall accuracy of the granular solver.

	In our implementation, the density and the velocity are updated by computing derivatives for several functions of density as well as the shear stress tensor. This is not the customary practice in granular codes. In \cite{Dunatunga2015Sep} and \cite{Yang2020May}, for example, the authors directly approximate the Cauchy stress tensor. This, however, requires components of the Cauchy stress tensor for the free surface and solid boundaries which are difficult to approximate accurately. In our approach, implementation of the boundary data is more straightforward.

	\item \textbf{Numerical Implementation of Boundary Conditions}
	Three types of boundary conditions are implemented in the granular flow code.
	\begin{itemize}
	\item Mirror particles are used for the {\bf solid boundary}. For every fluid particle within a prescribed distance to the boundary, a mirror particle is created and placed symmetrically across the local tangent plane. The mirror particle is assigned the same density of the corresponding fluid particle, the opposite component of the normal velocity, and zero-norm of the shear rate tensor. 
	
	\item Ghost particles are used for the {\bf free boundary}. If a particle is unable to find enough neighbors to calculate the differential operators, ghost particles are created to complete the neighborhood. Ghost particles are assigned states that model the corresponding fluid particle as a particle located on a free surface.  These include the continuity of pressure and velocity. The density of a ghost particle is slightly less than the minimum density as described in equation (\ref{eqn:EOS2}). Using these assumptions, the following expression for the shear rate tensor norm is calculated:
	\begin{equation}
	\left|\dot{\gamma}\right|_G = \left|\dot{\gamma}\right|_F \frac{\phi_c - \phi_G}{\phi_c - \phi_F}.
	\end{equation}
		
\item {\bf Periodic boundary} conditions periodically reflect states of Lagrangian particles between the left and right edges along one or two-dimensions in 3D. This capability was built using special functions of the p4est library.
\end{itemize}

	\item \textbf{Particle displacement.} Finally, the states are updated and the particles are moved to their new position in space, using the average of the old velocity and the new velocity. 
	\[
	\frac{x^{n+1} - x^n}{\Delta t } = \frac{1}{2}\left( u^n + u^{n+1} \right).
	\]

\end{enumerate}

\subsection{Particle data structure and parallelization algorithms} \label{sec:Octree}

The Lagrangian Particle code stores all the particle data in a single array. The elements
in the array contain all the information about the particles, including coordinates, physical states and some control variables. This makes maintenance and extension of the code easier and simplifies the communication of particle data among processes since the particle data is stored in adjacent memory blocks. 

The major octree algorithms such as building, refining and searching that significantly affect the code performance are parallelized using the p4est library. p4est \cite{Burstedde} enables a dynamic management of a collection of adaptive octrees on distributed memory supercomputers. It has the functionality of building, refining, coarsening, 2:1 balancing and partitioning on computational domains composed of multiple, connected, two-dimensional quadtrees or three-dimensional octrees, referred to as a forest of octrees. The granular flow model uses the same parallel structures and the communication / particle-redistribution algorithms as the earlier Lagrangian particle hydrodynamic code. Parallel algorithms in the hydrodynamic LP code are described in \cite{Samul2022}. p4est-based implementation of another Lagrangian particle code AP-Cloud (Adaptive Particle-in-Cloud) for optimal solutions of Vlasov-Poisson problems can be found in \cite{YuKumar2022}.

\subsection{Parallel scalability}

\begin{figure}[h!]
	\centering
	\subfigure[\label{fig:strongscaling}Strong scaling]{\includegraphics[width=0.7\linewidth]{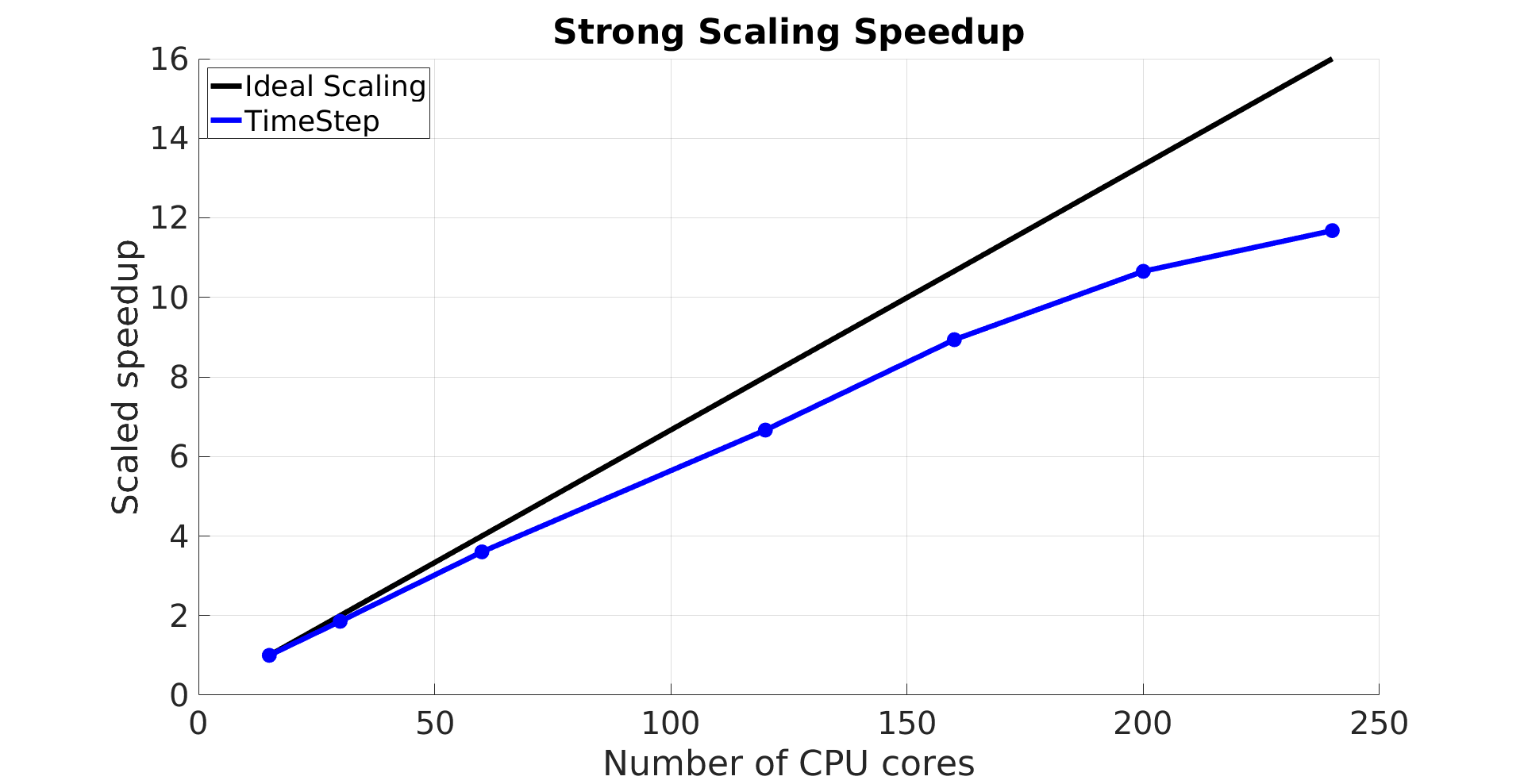}}
	\subfigure[\label{fig:weakscaling}Weak scaling]{\includegraphics[width=0.7\linewidth]{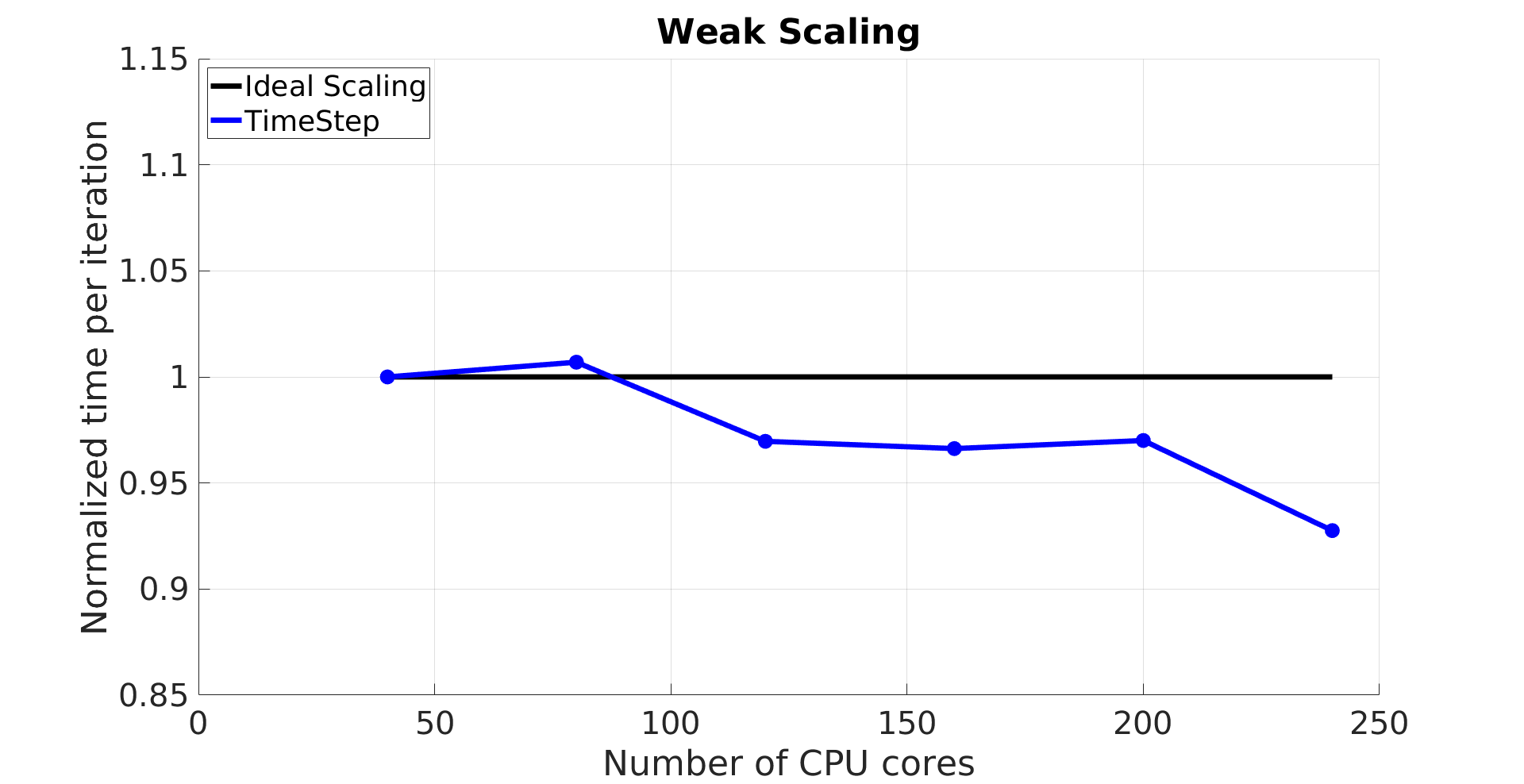}}
	\caption{Strong (a) and weak (b) scaling of the LP granular flow code.}
\end{figure}

We have studied parallel scalability of the code based on the 3D column collapse problem, described in the next section, using the Seawulf cluster located at Stony Brook University.  Strong scaling was tested by increasing the number of CPU cores while keeping the total number of computational particles constant (3,029,264 particles were used in the strong scaling tests).  Figure \ref{fig:strongscaling} shows strong scalability of the LP granular flow code.  

Weak scaling was measured by running the code with different number of CPU cores and using a proportionally scaled number of particles such that the number of particle  per core remained constant (25,000 particles per CPU core). The total number of Lagrangian particles varied from 1M for 40 CPU cores to 6M for 240 cores.  The results of weak scaling tests are demonstrated in Figure \ref{fig:weakscaling}.

The overall parallel scalability is very good. The weak scaling, which is more important for practical applications to large-size problems, is better compared to the strong one, as is typical for parallel codes. Achieving 97\% and 93\% of ideal speedup on 200 and 240 CPU cores, respectively, while running about 3M particles is sufficient for practical applications.

\section{Verification and validation of the granular flow code}\label{VandV}

In this section, results of numerical simulations and their comparison with experimental results are reported. We use the following granular flow model parameters, dependent on grain characteristics, that are consistent with experiments reported in \cite{Lube2004}:
\begin{itemize}
	\item Static angle of repose, $\mu_s$ = tan(30 $\deg$)
	\item Angle of Motion, $\mu_2$ = tan(50 $\deg$)
	\item Inertial number flow layer thickness, $I_0 = 2.65$	
	\item Grain diameter, $d = 0.32 mm$ 
	\item Intrinsic grain density, $\phi_s = 2600 kg/m^3$
\end{itemize}
For the function that relates the inertial number and the density (\ref{eqn:inertialdensity}), typical values of $\phi_c = 0.60$ and $\phi_{min} = 0.35$ are used. 

\subsection{Collapse of granular layer}

One of the key elements that numerical experiments of granular flow should be able to reproduce is compressibility. Given an initial bulk of granular material, not at critical density, the gravity should compresses the material up to the critical density.

\begin{figure}[h]
	\centering
	\includegraphics[width=.6\linewidth]{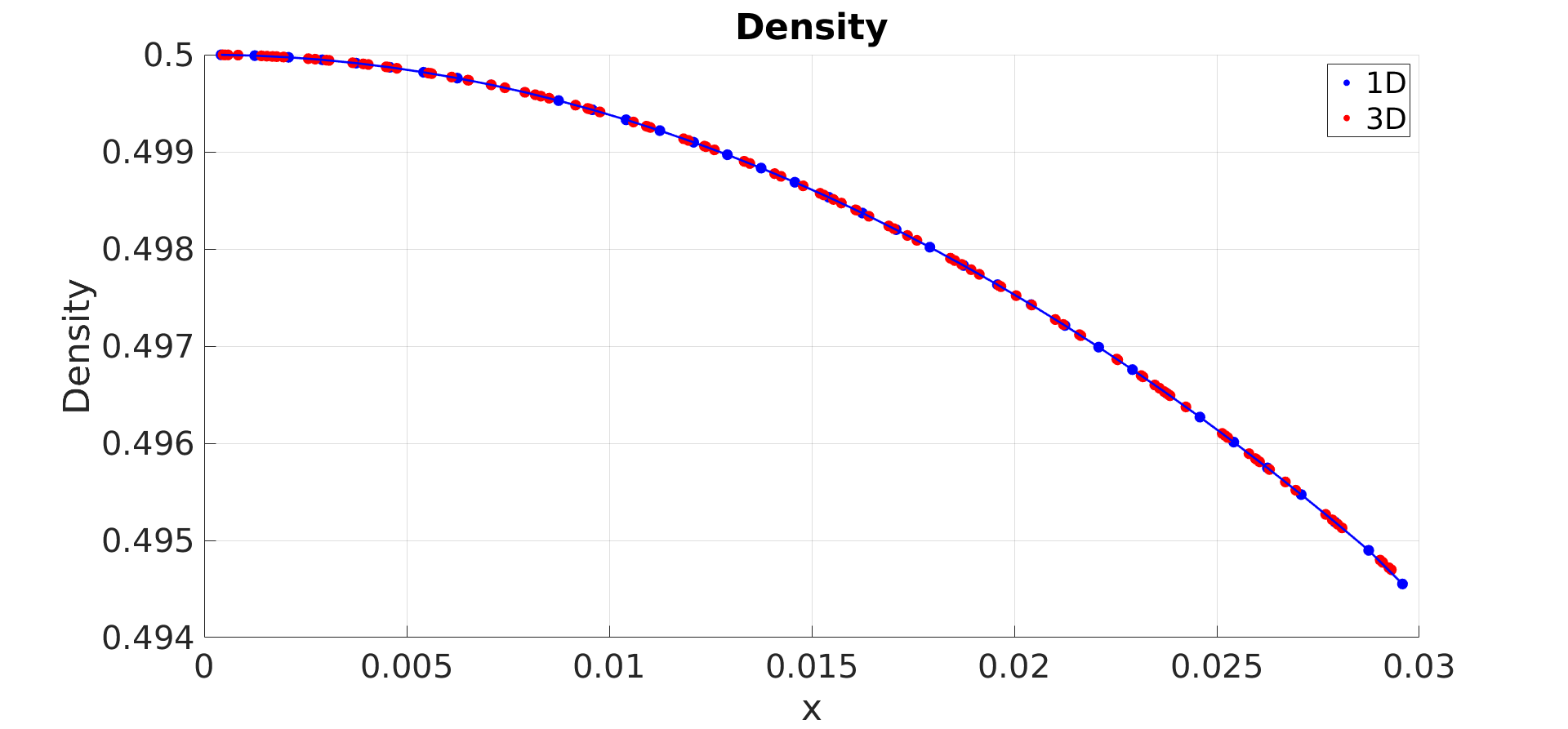}
	\caption{Initial density distribution for 1D and 3D compression simulations.}
	\label{fig:1D3DInitial}
\end{figure}

In order to assess the model capability to reproduce this property, two numerical experiments were performed. The first one consisted in the reduction of the granular system of equations to the 1D spacial dimension,  in which the compressibility depends only on the pressure term. A 1D column physically represents a 3D column of infinite transverse size. The same object was then modeled by the 3D LP granular flow code using periodic boundary conditions in both transverse directions. In both cases, a granular layer of 0.03m height  was initialized with the initial states of zero pressure, zero velocity, and a small initial gradient of density, as shown in Figure \ref{fig:1D3DInitial}.

The expected final height in this particular experiment is $0.025m$, and both the 3D code with periodic boundary conditions and the reduced 1D code gave virtually the same final states. In both cases, the final height of $0.02m$ was obtained.  The over-compression by 20\% suggests that the pressure relation given in (\ref{eqn:EOS2}) may not be the most accurate when the density is in the lower part of the interval $[ \phi_{min} , \phi_{max} ]$ of the inertial regime. Moreover, for all other numerical tests, the resulting final height was always slightly less than the expected final height, with the over-compression error decreasing as the initial density approaches the critical density $\phi_c$. The good agreement between 1D and 3D code suggests that the verification test of the solvers was successful, and the discrepancy wth the experiment is dependent on the details of the closure model.

\subsection{Axisymmetric collapsing column}

The study of a collapsing axisymmetric column is a common benchmark for the validation of granular codes. Work \cite{Lube2004} reports experimental results on granular columns, initially confined by a cylindrical tube, and then set into motion by rapidly lifting the tube. By systematic experimental studies, the author of \cite{Lube2004} identified three regimes depending on the aspect ratio $(a = h_i/r_i)$ between the initial height and radius of the column. The final radius $r_{\infty}$ for the first two regimes is given by
\begin{equation}\label{eqn:fitting1}
r_{\infty} = 
\begin{cases}
r_i \left(  1+1.24a\right),  &\quad\text{if } 0 < a<1.7 \\
r_i \left(  1+1.6 a^{1/2} \right),  &\quad\text{if } 1.7<a,<10. \\
\end{cases}
\end{equation}  
The final radius as well as the final height of the resulting cone depend on two processes, the initial collapsing and the subsequent avalanching. The resulting power laws for the final height are
\begin{equation}\label{eqn:fitting2}
h_{\infty} = 
\begin{cases}
a r_i,  &\quad\text{if } 0<a<1.7, \\
0.88 r_i  a^{1/6},   &\quad\text{if } 1.7<a < 10.\\
\end{cases}
\end{equation}
The time needed to reach $r_{\infty}$ depends only on the initial height and is given as 
\begin{equation}\label{eqn:fitting3}
t_{\infty} = 3 \sqrt{\frac{h_i}{g}}.
\end{equation}

 For flows arising from columns with aspect ratios less than 1.7, although the final radius and final height are given by equations \ref{eqn:fitting1} and \ref{eqn:fitting2}, \cite{Lube2004} makes further divisions based on the evolution of the flow to steady state. Specifically, for columns with aspect ratio less than 0.74, a single circular discontinuity forms in the free surfaces, separating an outer avalanching region from an inner non-deformed cylindrical section. 

 \subsubsection{Columns with aspect ratio $a < 1.7$}  
  
\begin{figure}[h!]
 	\centering
 	\subfigure[\label{fig:60travel1}Transversal Velocity]{\includegraphics[width=.7\textwidth]{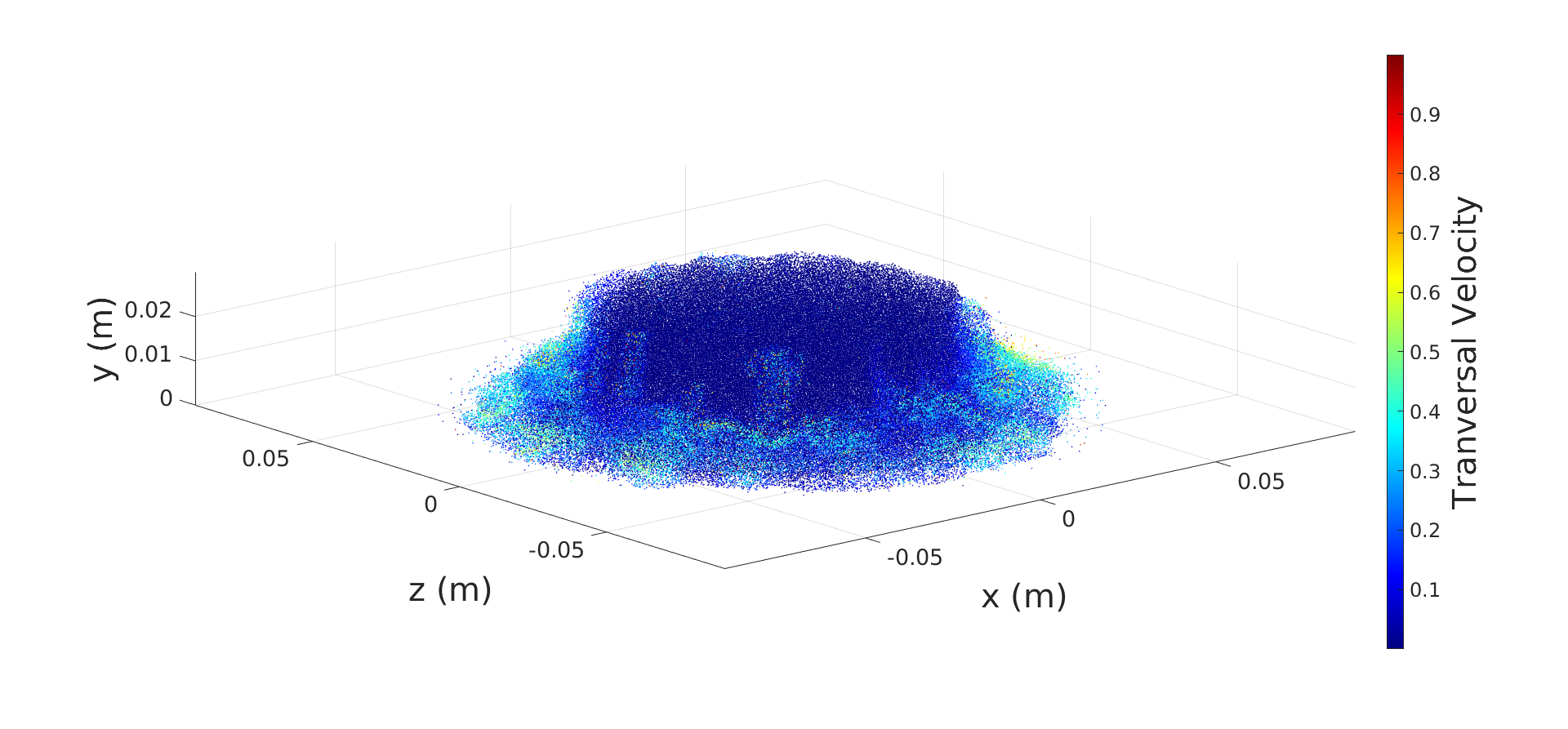}}
  	\subfigure[\label{fig:60density1}Density slice]{\includegraphics[width=.7\textwidth]{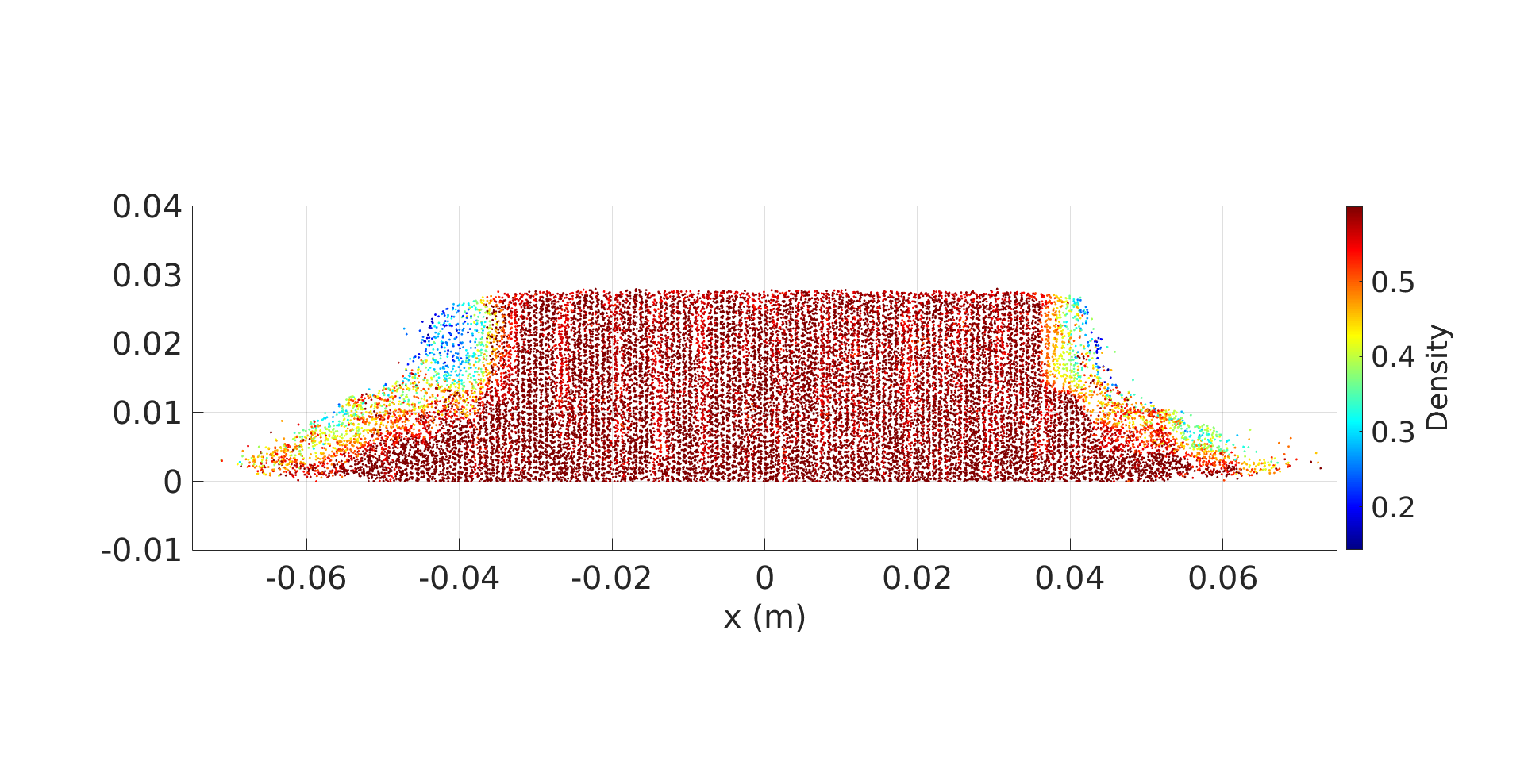}}
 	\caption{Transverse velocity (a) and density slice (b) at $t = 0.07s$ for collapsing column with the initial aspect ratio $a=0.60$.}
 \end{figure}

 The first numerical simulation  was performed for a column with the aspect ratio $a = 0.6$. Figure \ref{fig:60travel1} depicts the transverse velocity at intermediate time 0.07s and demonstrates the avalanching process and Figure \ref{fig:60density1} shows the transverse distribution of density at $t=0.07$s. We observe that the density remains mostly constant, except in the outermost layers where the avalanching occurs. 
 
\begin{figure}[h!]
 	\centering
 	\includegraphics[width=.7\textwidth]{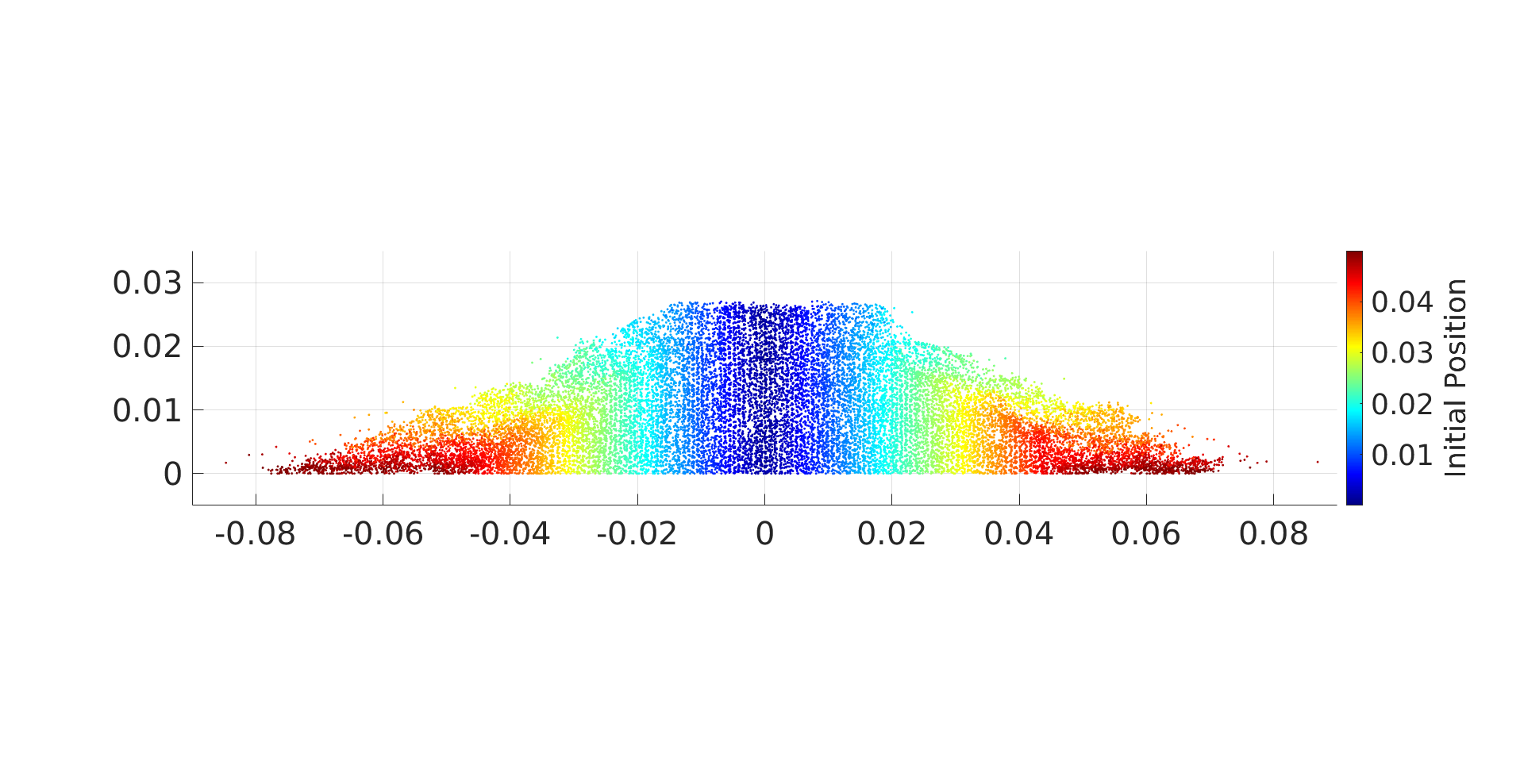}
 	\caption{Final layering at $t = 0.17s$ for the column with $a=0.60$}
 	\label{fig:60initR}
 \end{figure}

 A defining characteristic of the collapsing column dynamics is the final layering of particles with respect to the initial configuration. Figure \ref{fig:60initR} shows that the final layering has very good agreement with the layering described in \cite{Lube2004}. It is also worth noting that the inner non-deformed cylindrical section is observed in Figure \ref{fig:60initR}.
 
 Based on the empirical expressions, the expected height of the cone at $t = 0.17s $ is $ h_{\infty} = 0.03m$ and the final radius is $r_{\infty} = 0.087m$. The final height obtained from the numerical simulation is $0.027m$, the final radius is about $0.085m$. Both quantities agree with the experiment within an acceptable margin.
 
\begin{figure}[h!]
 	\centering
 	\subfigure[\label{fig:120t1}Early time]{\includegraphics[width=0.7\linewidth]{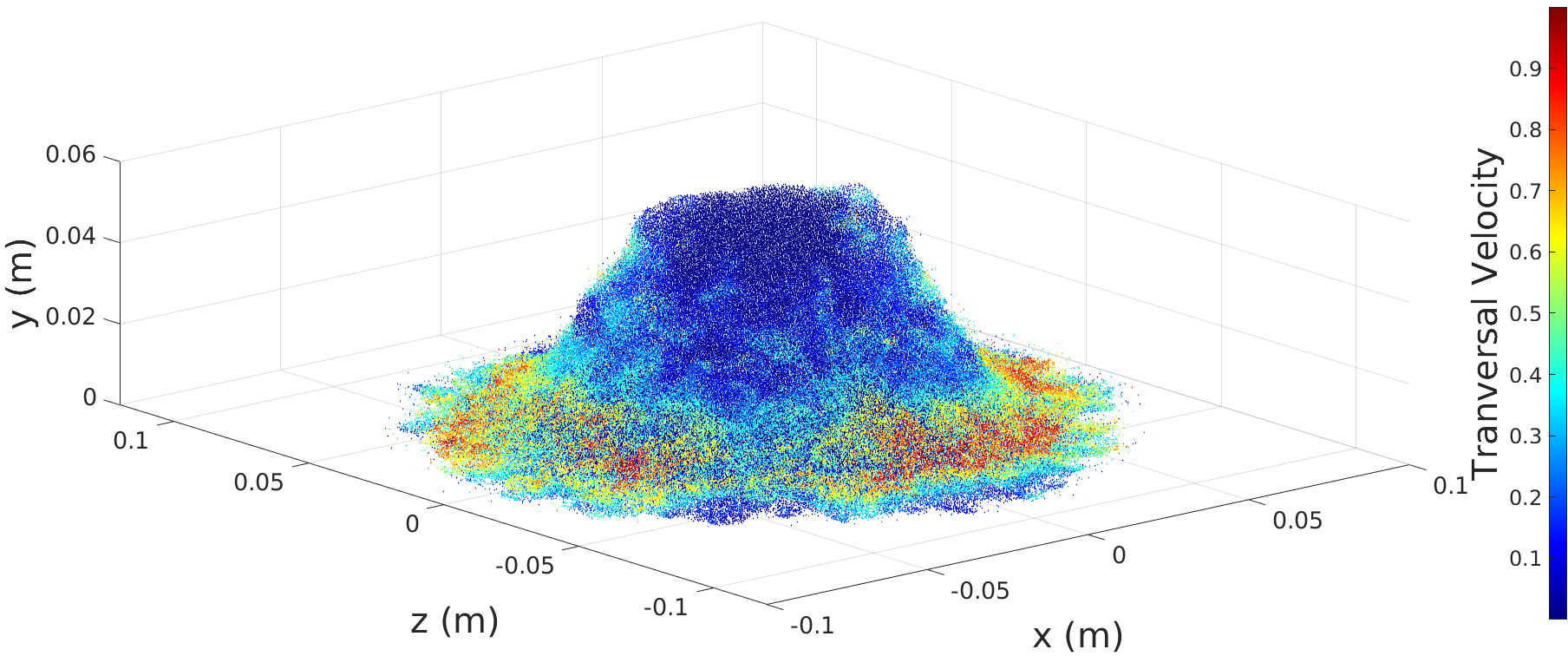}}
 	\subfigure[\label{fig:120t2}Late time]{\includegraphics[width=0.7\linewidth]{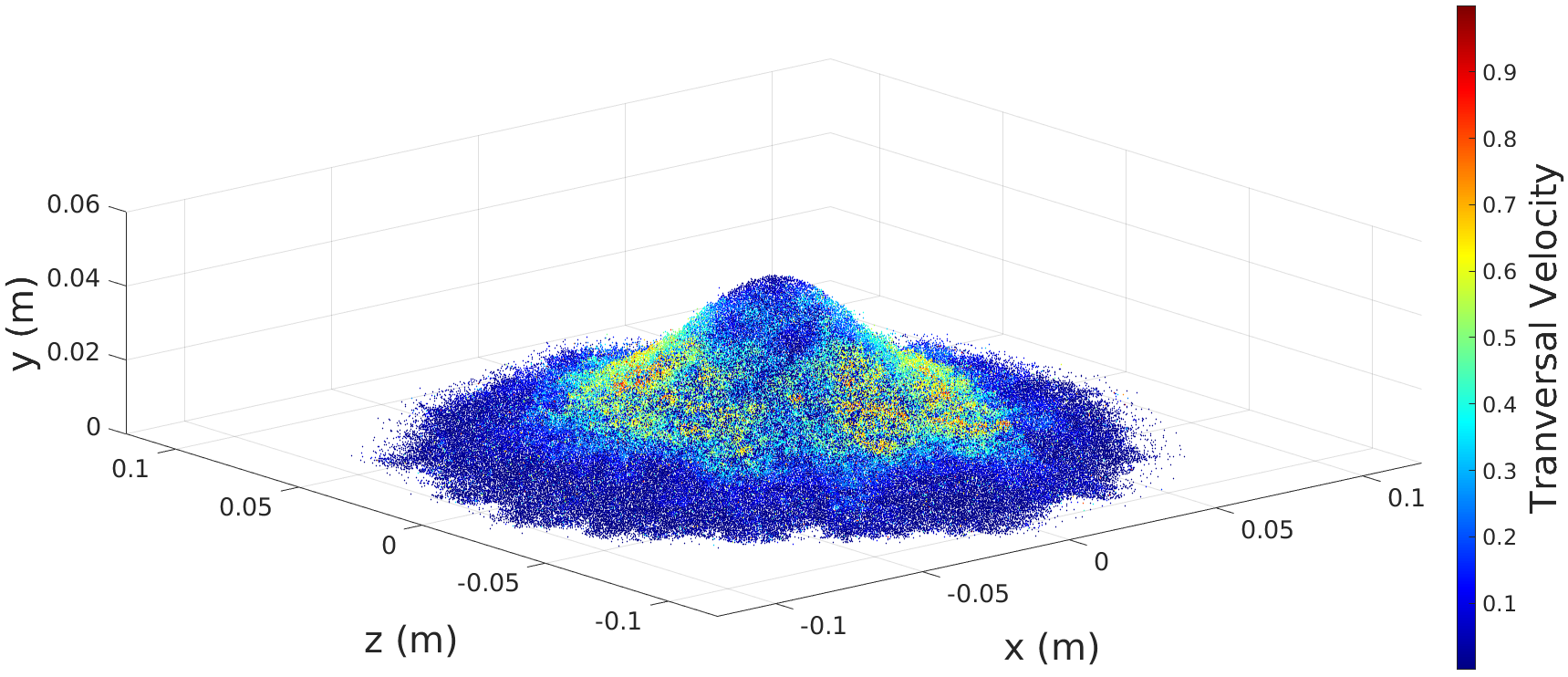}}
 	\caption{Transverse velocity at early time(a) and after reaching the final radius (b) for collapsing column with the initial aspect ratio $a = 1.2$ }
 \end{figure}

 For flows with aspect ratio greater than 0.74 the evolution is more complex. Work \cite{Lube2004} describes the formation of two circular discontinuities in the free surface that divide the column into three concentric regions. The granular material in the outermost region starts flowing first, than the material in the middle region, as can be observed in figure \ref{fig:120t1}. The particles in the outermost region are the first to stop moving, then the slope of the free surface increases and eventually reaches the angle of repose. This can be observed in figure \ref{fig:120t2}. 
 
\begin{figure}[h!]
 	\centering
 	\subfigure[\label{fig:90t21}$a = 0.9$]{\includegraphics[width=0.7\linewidth]{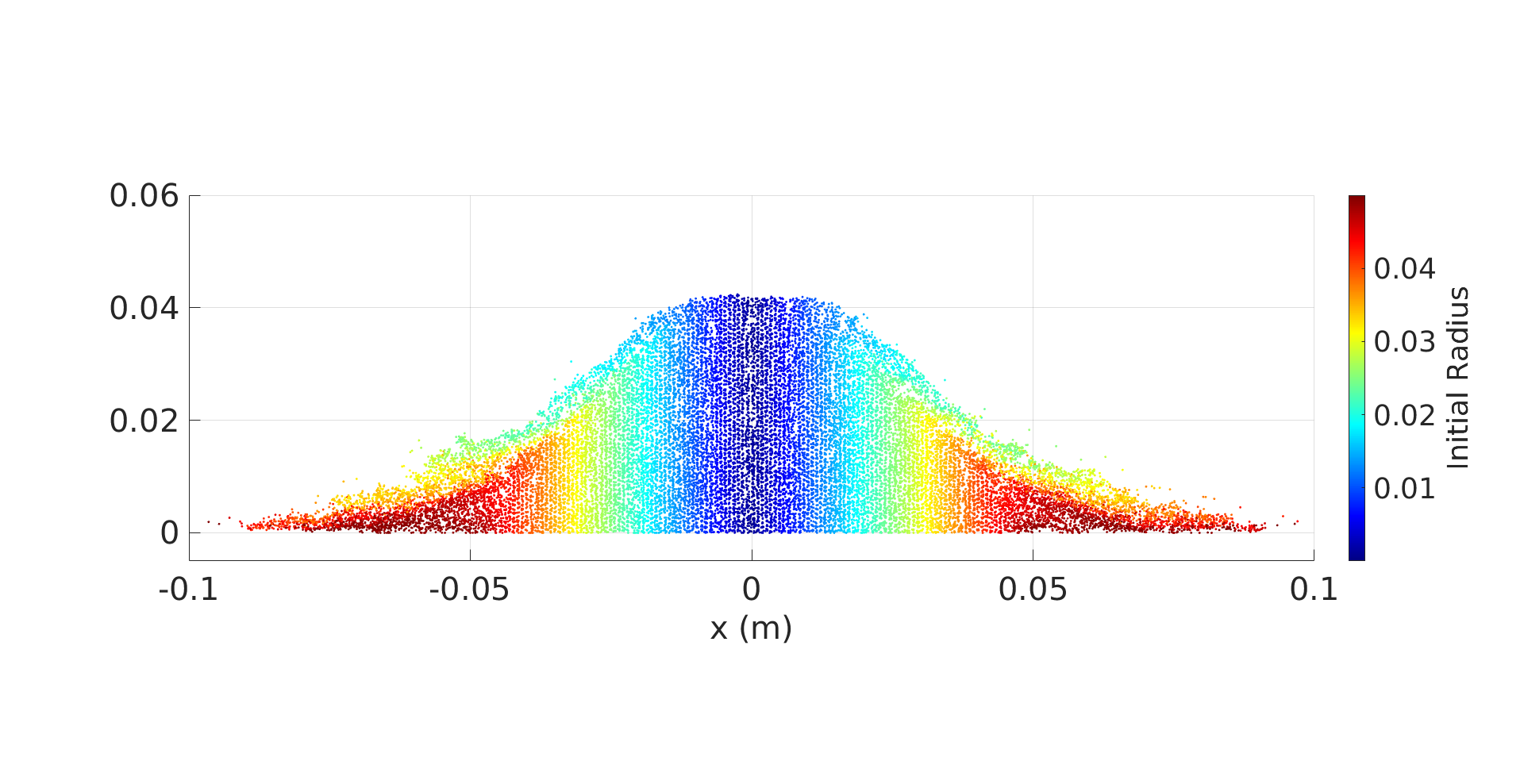}}
 	\subfigure[\label{fig:120rfinal}$a = 1.2$]{\includegraphics[width=0.7\linewidth]{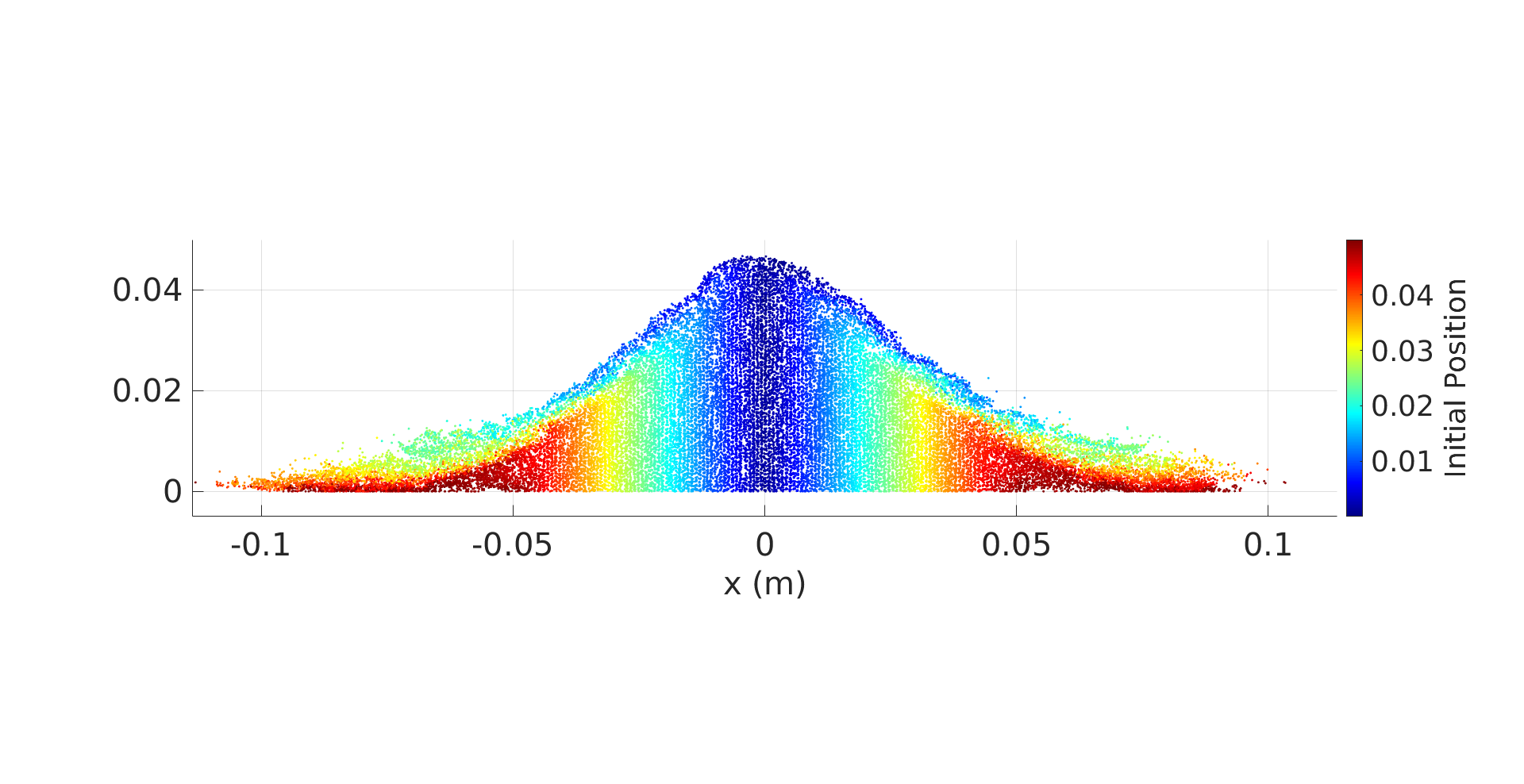}}
 	\caption{Distribution of granular layers  for collapsed column with aspect ratio $a=0.9$ (a) and $a = 1.2$ (b).}
 \end{figure}

 If the aspect ratio is less than 1, an inner cylindrical region stays constant through all the process. But for aspect ratios greater than one, the avalanching process disrupts the inner cylindrical region and produces a conical pile as observed in Figures \ref{fig:90t21} and \ref{fig:120rfinal}. Simulation results are in good agreement with experimental images of \cite{Lube2004}.
 
 \subsubsection{Columns with aspect ratio $a > 1.7$}  
 
\begin{figure}[h!]
 	\centering
 	\subfigure[\label{fig:200transinitradius2}Intermediate time]{\includegraphics[width=0.7\linewidth]{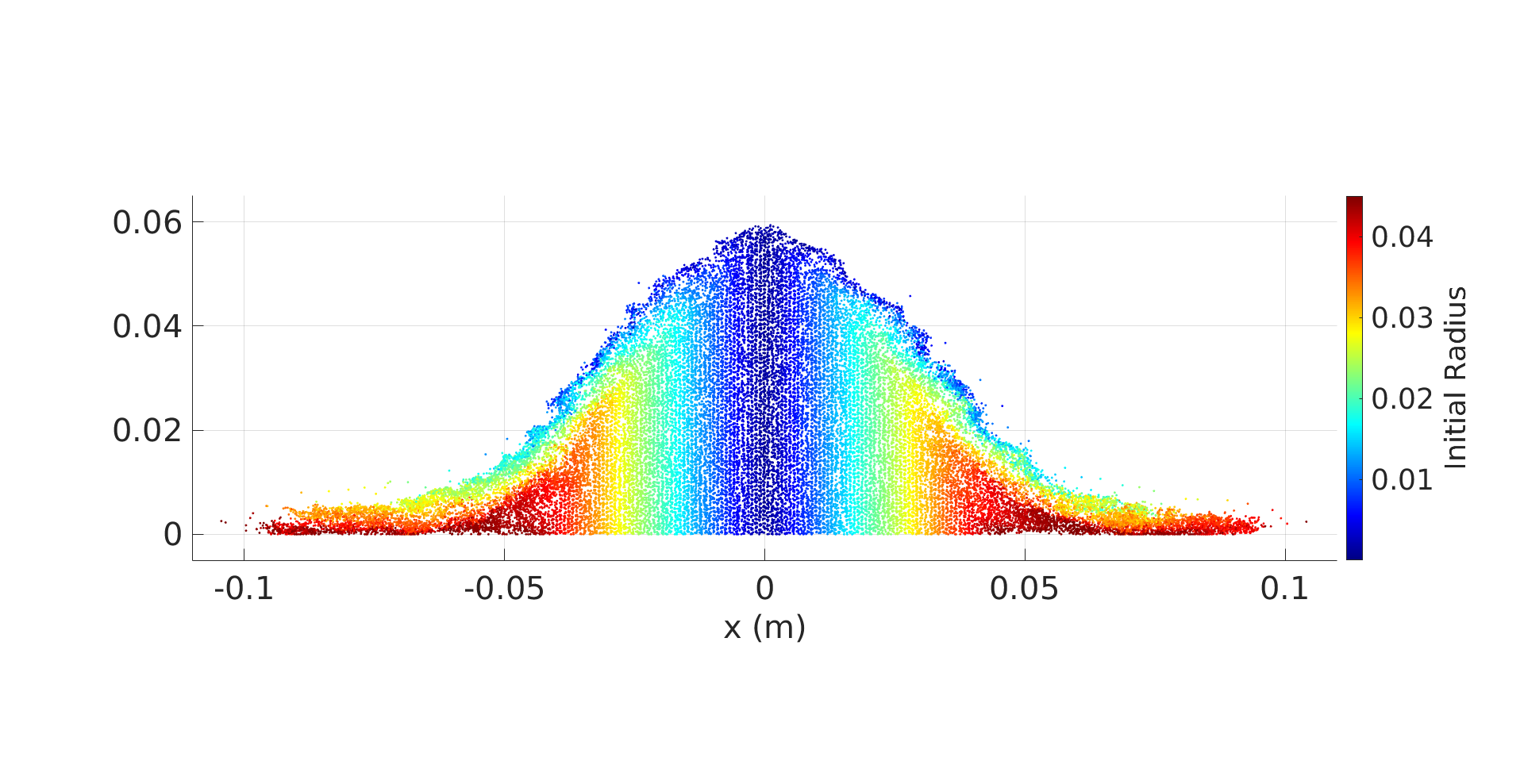}}
 	\subfigure[\label{fig:200transinitradius4}Final time]{\includegraphics[width=0.7\linewidth]{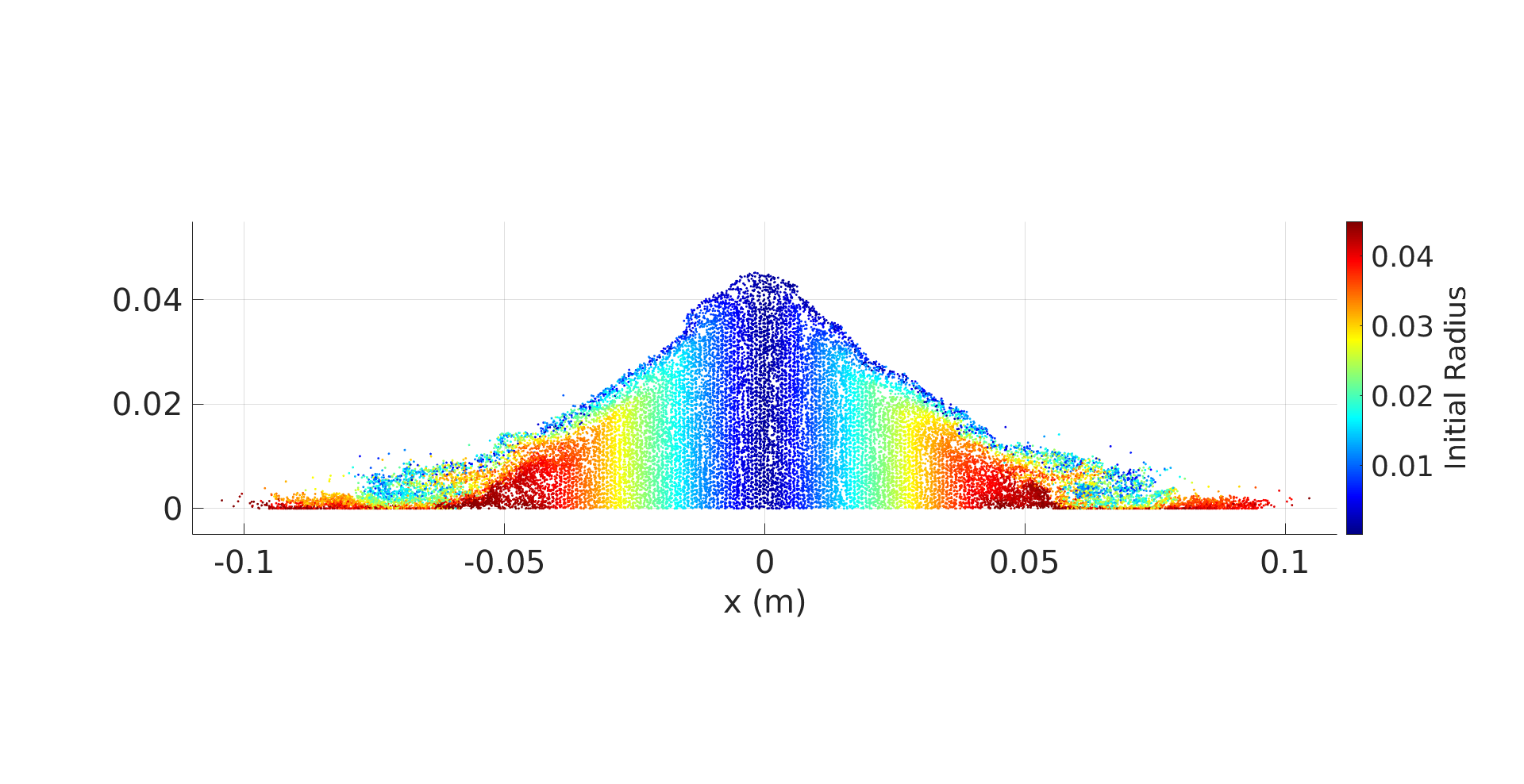}}
 	\caption{Slice of granular material showing the intermediate (a) and final (b) layering of particles based on their initial radius.}
 \end{figure}

 For granular columns with the aspect ratio greater than 1.7, the final shape of the heap is dominated by a process of collapsing and avalanching. The particles in the top layers roll down the slope that is formed by the particles in the base moving outwards, as can be observed in Figure \ref{fig:200transinitradius2}. The final geometry shown in Figure \ref{fig:200transinitradius4} is qualitatively similar. These images also demonstrate the distinctive characteristic of the collapsing column, namely that a cone of $45^{\circ}$ remains motionless through all the process. The final layering obtained from the simulation is similar to the schematic that summarizes experimental observations (see Figure 5 in \cite{Lube2004}). 
 
 For large aspect ratios,  the final radii from simulation were consistently smaller by about 20\% compared to the experimental fit estimates by (\ref{eqn:fitting2}).  This discrepancy is perhaps caused by the closure model. For the inertial regime when $\phi_{min} < \phi < \phi_{max}$, $ P \sim \left|\dot{\gamma}\right|^2 $ in our closure model (\ref{muofphi}) that is consistent with the Bagnold scaling \cite{Bagnold}. However, \cite{Chialvo2012} reports that this quadratic dependence turns into $P \sim \left|\dot{\gamma}\right|^{\frac{1}{2}}$ in the lower part of $( \phi_{min} , \phi_{max}) $.

 \subsection{Transition to the gas-like regime} \label{sec:GasLike}
 
 As the density of the granular material approaches the minimum density $\phi_{min}$, grains lose contact with each other and the pressure approaches zero. In this regime, the movement of the material is controlled only by the grain inertia. In order to assess the capabilities of the model that supports the transition into the gas-like regime, we performed simulations of a collapsing column at a small distance from the edge of a horizontal support plane raised above the ground. Expanding initially in all directions, granular material goes into free fall on one side of the granular column. 
 
\begin{figure}[h!]
 	\centering
 	\subfigure[\label{fig:f1slope}Slope difference]{\includegraphics[width=.7\linewidth]{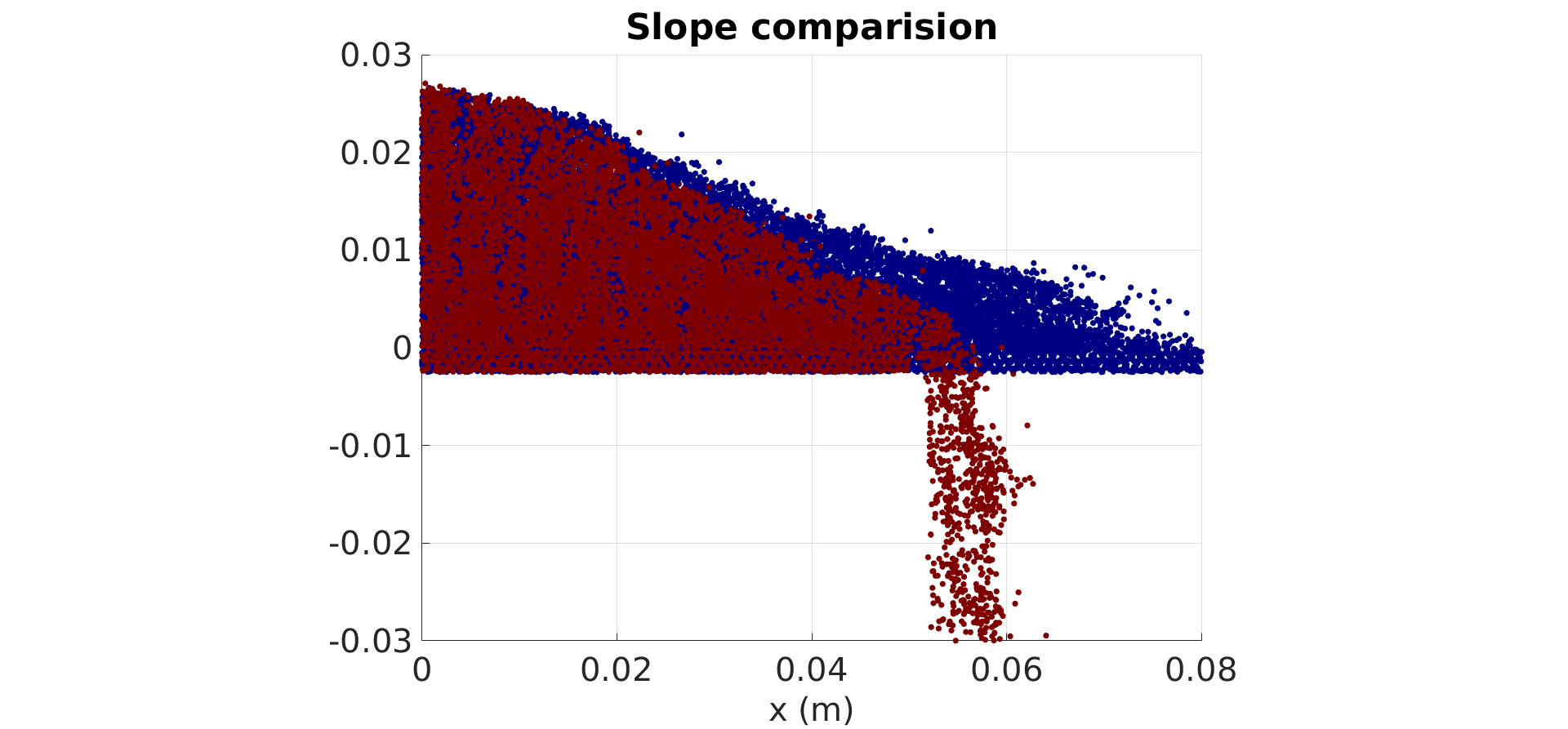}}
 	\subfigure[\label{fig:FallingTransvelocity}Transverse velocity]{\includegraphics[width=.7\linewidth]{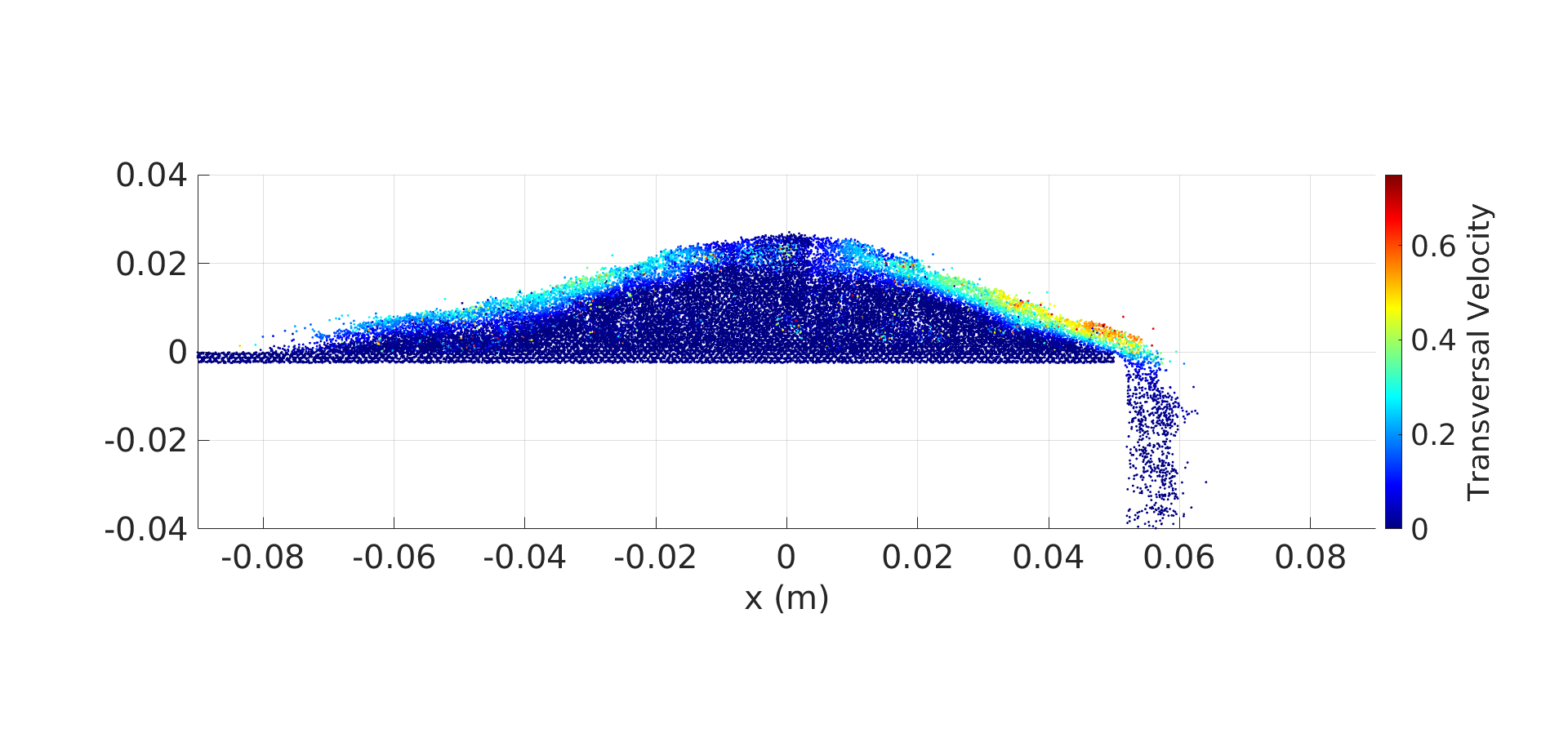}}
 	\caption{(a) Difference in slopes for collapsing column with and without solid boundary edge. (b) Transverse velocity in the collapsing at the edge of the solid boundary.}
 	
 \end{figure}

 In the regions with full solid boundary, the column collapses as usual into a grain heap with the characteristic cone shape. The slope of this heap is  defined by the avalanching process that deposits layers of grains on top of each other. In the direction where the solid boundary is absent, the particles start an abrupt free fall and the heap slope changes (Figure \ref{fig:f1slope}).
 Due to smaller friction experienced by granular layers on the short slope during the avalanche process, the transverse velocity is higher on the side of the free fall as shown in Figure \ref{fig:FallingTransvelocity}. 
 
\section{Conclusion and future work}

A numerical model and parallel software for 3D simulations of granular flows have been developed based on the Lagrangian particle (LP) method for compressible hydrodynamic flows that resolves free surface and multiphase problems. The granular flow model implements continuum equations with a $\mu(I)$-rheology closure that is capable of describing two-directional transitions of the flow between various regimes characterized by solid-, liquid-, and gas-like features. 

The Lagrangian particle-based discretization of spatial derivatives uses local least squares polynomial fitting or generalized finite differences on particle stencils that ensures numerical convergence to the prescribed order. The continuum adaptivity property of Lagrangian particles is ideally suitable to the simulation of flows with large density changes. The current discretization is second-order accurate in space and first-order accurate in time. Because of the need to use very small time step due to rapid density changes, the 1st order time discretization does not significantly reduce accuracy. A second-order accurate time discretization will be implemented in the future.  

The Lagrangian particle code uses octrees as the particle data structure and the octree construction and search routines provide optimal particle neighborhoods for numerical stencils. The major octree algorithms such as building, refining and searching that significantly affect the code performance are parallelized using the p4est library. p4est (Parallel Forest of K-trees) enables a dynamic management of a collection of adaptive octrees on distributed memory supercomputers. It has the functionality of building, refining, coarsening, 2:1 balancing and partitioning on computational domains composed of multiple, connected, two-dimensional quadtrees or three-dimensional octrees, referred to as a forest of octrees. The granular flow model uses the same parallel structures and the communication / particle-redistribution algorithms as the earlier Lagrangian particle hydrodynamic code.

The LP granular flow code has been validated by comparing 3D simulations to experimental data on the collapse of granular columns. Numerical simulations are in good agreement with experiments on the time-scale dynamics, the final shape of the granulas material heap, and the layering of grains depending on their initial location. The code has reproduced the major experimental features for granular columns with various values of the aspect ratio defined as the initial column height divided by the initial radius. For column with the largest aspect ratio, the simulated final radius was about 20\% smaller compared to experiments. We believe that the pressure relation on the closure model could be responsible for this error. We have also demonstrated numerical simulation of a granular flow undergoing a change of the flow regime with grains losing contacts at reduced density using an example of collapsing column at the edge of a horizontal support plane. 

Future developments will include improvements of both numerical algorithms and the closure model. In the area of algorithms, we will implement second order time stepping. The whole algorithm can also be easily upgraded to higher degree of overall accuracy by using higher order polynomial fittings for spatial derivatives and a Runge-Kutta time integration within the method if lines. A locally adaptive time stepping would be greatly beneficial for problems with changing flow regimes. We will also investigate other relations within the closure model and perform parametric and sensitivity studies.

{\bf Acknowledgments}
This work was supported in part by the U.S. Department of Energy grant No. DE-SC0014043.



\bibliographystyle{elsarticle-num} 
\bibliography{biblio}


\end{document}